\documentclass[twocolumn,prl,aps,showpacs,floatfix]{revtex4}

\usepackage{graphicx}%

\usepackage{dcolumn}
\usepackage{amsmath}

\newcommand{\ice}[1]{\relax}

\newcommand{\unl}{\underline}

\newcommand{\as}{a_s}

\newcommand{\beq}{\begin{equation}}

\newcommand{\ba}{\begin{array}}

\newcommand{\ea}{\end{array}}

\newcommand{\eeq}{\end{equation}}

\newcommand{\bea}{\begin{eqnarray}}

\newcommand{\eea}{\end{eqnarray}}

\newcommand{\ep}{\epsilon}

\newcommand{\dd}{{\rm d}}

\begin{document}

\title{
{
 \vspace*{-16mm}
\centerline{\normalsize\hfill  SFB/CPP-08-04}
\centerline{\normalsize\hfill  TTP08-01    }
\baselineskip 11pt
}
{}

\vspace{1mm}
  Hadronic $Z$- and $\tau$-Decays in Order $\alpha_s^4$

 }     

\author{P.~A.~Baikov}
\affiliation{Institute of Nuclear Physics, Moscow State University,
Moscow~119899, Russia
        }
\author{K.~G.~Chetyrkin}\thanks{{\small Permanent address:
Institute for Nuclear Research, Russian Academy of Sciences,
 Moscow 117312, Russia}}

\author{J.~H.~K\"uhn}
\affiliation{Institut f\"ur Theoretische Teilchenphysik,
  Universit\"at Karlsruhe, D-76128 Karlsruhe, Germany}

\begin{abstract}
\noindent

Using recently developed methods for the evaluation of five-loop
amplitudes in perturbative QCD, corrections of order $\alpha_s^4$ for
the non-singlet part of the cross section for electron-positron
annihilation into hadrons and for the decay rates of the $Z$-boson and
the $\tau$-lepton into hadrons are evaluated.  The new terms lead to a
significant stabilization of the perturbative series, to a reduction
of the theory uncertainly in the strong coupling constant $\alpha_s$,
as extracted from these measurements, and to a small shift of the
central value, moving the two central values closer together.  The
agreement between two values of $\alpha_s$ measured at vastly
different energies constitutes a striking test of asymptotic freedom.
Combining the results from $Z$ and $\tau$ decays we find
$\alpha_s(M_Z)=0.1198 \pm 0.0015$ as one of the most precise and
presently only result for the strong coupling constant in order  $\alpha_s^4$.

\end{abstract}


\maketitle


The strong coupling constant $\alpha_s$ is one of the three
fundamental gauge couplings constants of the Standard Model (SM) of
particle physics.  Its precise determination is one of the most
important aims of particle physics. Experiments at different energies
allow to test the predictions for its energy dependence based on the
renormalization group equations, the comparison of the results
obtained from different processes leads to critical tests of the
theory and potentially to the discovery of physics beyond the Standard
Model. Last but not least, the convergence of the three gauge coupling
constants related by  SU(3)xSU(2)xU(1) to a common value, after
evolving them to high energies, allows us  to draw conclusions about the
possibility of embedding the SM in the framework of a Grand Unified
Theory.

One of the most precise and theoretically safe determination 
of $\alpha_s$ is based on
measurements of the cross section for electron-positron annihilation
into hadrons. These have been performed in the low-energy region between 2~GeV
and 10~GeV and, in particular, at and around the $Z$ resonance at 
91.2~GeV. 
Conceptually closely related is the measurement of the semileptonic
decay rate of the $\tau$-lepton, leading to a determination of $\alpha_s$ at
a scale below 2 GeV.


From the theoretical side, in the framework of perturbative QCD, these rates
and cross sections are evaluated as inclusive rates into massless quarks and gluons 
\cite{ChKK:Report:1996,Davier:2005xq}.
(Power suppressed  mass effects are  well under control for
$e^+e^-$-annihilation,  both at low energies and around the $Z$ resonance, 
and for $\tau$ decays 
\protect\cite{Chetyrkin:1996ii,Chetyrkin:1996cf,ChetKuhn90,ChK:mq4as2,ChKH:mq4as3,Baikov:2004ku}, 
and the same applies to mixed QCD and electroweak corrections \cite{Czarnecki:1996ei,Harlander:1997zb}).

The ratio 
$R(s)\equiv \sigma(e^+e^-\to {\rm hadrons}) / \sigma(e^+e^-\to\mu^+\mu^-)$ 
is 
expressed through the absorptive part of the 
correlator of the electromagnetic current $j_\mu$:
\begin{eqnarray}
R(s) &=& 12\pi\, \mathrm{Im}\, \Pi(-s - i\epsilon)
,
\\
3\, Q^2\, \Pi(Q^2) &=& 
i\int\dd^4 xe^{iq\cdot x}\langle 0|{\rm  T}j_\mu(x)j^\mu(0)|0\rangle
{},
\end{eqnarray}  
with $Q^2 = -q^2$. 
It is also convenient to introduce  the Adler function as
\beq
\begin{array}{c}
\nonumber
{D}(Q^2) =  -12\, \pi^2
Q^2\, \frac{\mathrm{d}}{\mathrm{d} Q^2} \Pi(Q^2)
=
\int_0^\infty \frac{ Q^2\ R(s) d s }{(s+Q^2)^2}
\nonumber
{},
\\
R(s) = \ice{\frac{1}{2\pi i} \int_{-s -i\ep}^{-s +i\ep} \frac{d\, Q^2}{Q^2}=}
D(s) - \pi^2 \beta_0^2 \{ \frac{ d_1}{3} a_s^3 + (d_2 + \frac{5}{6\,\beta_0 }d_1\,\beta_1)\,a_s^4\} + \dots
{}\ .
\end{array}
\eeq
We define  the perturbative  expansions 
\beq
{D}(Q^2) = \sum_{i=0}^{\infty} \  {d}_i a_s^i(Q^2),  \ \
{R}(s) =  \sum_{i=0}^{\infty} \  {r}_i a_s^i(s)
{},  
\eeq
where $a_s \equiv \alpha_s/\pi$ and  
the normalization scale  is set to  $\mu^2=Q^2$ or to $\mu^2= s$
for the Euclidian and Minkowskian functions respectively.
The results
for generic values of $\mu$ can be easily recovered with  standard
RG techniques.

Note that the first three terms of the perturbative series for $D$ and $R$ 
coincide. Starting from $r_3$, terms proportional $\pi^2$ arise which
can be predicted from those of lower order.
It has been speculated that these ``$\pi^2$-terms'', also  called ``kinematical  terms'', 
 might constitute a major
part of the full higher order corrections (see, e.g. \cite{Kataev:1995vh,Shirkov:2000qv} 
and references therein); 
however, the validity of this
hypothesis can only be established by the full calculation. Indeed, for the
scalar correlator this assumption has been shown to fail
\cite{Baikov:2005rw}.


For the vector correlator the terms of order $a_s^2$ and $a_s^3$ have been
evaluated nearly thirty and about fifteen years ago 
\cite{Chetyrkin:1979bj,Gorishnii:1991vf,Surguladze:1991tg}, respectively. 
The $a_s^4$ corrections are conveniently classified according to their 
power of $n_f$, with $n_f$ denoting the number of light quarks.  
The $a_s^4 n_f^3$ term is part of the ``renormalon chain'', the evaluation of
the next term, of order $a_s^4 n_f^2$, was a test case for the techniques used
extensively in this paper and, furthermore, led to useful
insights into the structure of the perturbative series already 
\cite{ChBK:vv:as4nf2}.

The complete five-loop calculation requires the evaluation of 
about twenty thousand diagrams (we have  used QGRAF \cite{Nogueira:1991ex} for
their automatic generation). 
Using ``infrared rearrangement''
\cite{Vladimirov:1980zm}, the
$R^*$ operation \cite{ChS:R*} and the prescriptions formulated in
\cite{gvvq} to algorithmically resolve the necessary combinatorics, it
is possible to express the absorptive part of the five-loop diagrams
in terms of four-loop massless propagator integrals.

These integrals can be reduced to a sum of 28 master integrals
with rational functions of the space-time dimension $D$ as coefficients.
The latter ones were fully reconstructed after evaluating sufficiently
many terms of the $1/D$ expansion \cite{Baikov:2005nv} of their
representation proposed in \cite{Baikov:tadpoles:96}. This direct and
largely automatic procedure required enormous computing resources and
was performed using a parallel version \cite{Tentyukov:2004hz} of FORM
\cite{Vermaseren:2000nd}.

In this paper we present the results for the so-called ``non-singlet''
diagrams. 
These are sufficient for a complete description of
$\tau$-decays. For $e^+e^-$ annihilation through a virtual photon 
they correspond to the dominant terms
proportional $\sum_i Q_i^2$. The singlet contributions proportional
$(\sum_i Q_i)^2$  arise for the first time in ${\cal O}(\alpha_s^3)$. They 
are known to be small, and will be evaluated at a later
point. Similar comments apply to the singlet contributions in $Z$ decays.

The analytic result for the five loop term in  the Adler function is given  by
(we suppress the trivial factor $3\,\sum_f Q^2_f$
throughout)
\begin{gather}
\textstyle
\label{d4}
d_4 = {} 
 n_f^3
\left[
-\frac{6131}{5832}
+\frac{203}{324}  \,\zeta_{3}
+\frac{5}{18}  \,\zeta_{5}
\right]
\\
\textstyle
{+} \, n_f^2
\left[
\frac{1045381}{15552}
-\frac{40655}{864}  \,\zeta_{3}
+\frac{5}{6}  \,\zeta_3^2
-\frac{260}{27}  \,\zeta_{5}
\right]
\nonumber\\
\textstyle
{+} \, n_f
\left[
-\frac{13044007}{10368}
+\frac{12205}{12}  \,\zeta_{3}
-55  \,\zeta_3^2
+\frac{29675}{432}  \,\zeta_{5}
+\frac{665}{72}  \,\zeta_{7}
\right]
\textstyle
\nonumber
\\
\textstyle
{+}
\frac{144939499}{20736}
-\frac{5693495}{864}  \,\zeta_{3}
+\frac{5445}{8}  \,\zeta_3^2
+\frac{65945}{288}  \,\zeta_{5}
-\frac{7315}{48}  \,\zeta_{7}
{}.
\nonumber
\end{gather}

The  knowledge of $d_4$ leads straightforwardly  to  $R$ at order $\alpha_s^4$, 
for  brevity given below  in  numerical form:
\bea
R &=& 1  + a_s +  (1.9857 - 0.1152\, n_f)\, a_s^2 
\nonumber
\\
&+&
(-6.63694 - 1.20013 n_f - 0.00518 n_f^2 ) \, a_s^3
\label{R_numerical_nl}
\\
\nonumber
&+&(-156.61 + 18.77\, n_f - 0.7974\, n_f^2  + 0.0215\,  n_f^3 ) \, a_s^4
{}.
\eea
It is also instructive to explicitly display  the genuine five-loop contributions to  $d_4$
(underlined in (\ref{r_3_separated},\ref{r_4_separated}) ) and  the  ``kinematical'' terms
originating from the analytic continuation:
\bea
r_3 &=& \unl{18.2} - 24.9  + (\unl{-4.22} + 3.02)\, n_f 
\label{r_3_separated}
\\
&{+}& (\unl{-0.086} + 0.091)\, n_f^2
\nonumber
{},
\\
r_4 &=& \unl{135.8} - 292.4  + (\unl{-34.4} + 53.2)\, n_f
\label{r_4_separated} 
\\
&+&(\unl{1.88} - 2.67)\, n_f^2 + (\unl{-0.010} + 0.032)\, n_f^3
\nonumber
{}.
\eea

Since it will presumably take a long time until the next term of the
perturbative series will be evaluated,
it is of interest to investigate
the predictive power of various optimization schemes empirically. Using the
principles of ``Fastest Apparent Convergence'' (FAC)  \cite{Grunberg:1984fw} 
or of ``Minimal
Sensitivity'' (PMS) \cite{Stevenson:1981vj}, which happen to coincide in this order, the central  values of 
the  predictions
\cite{Kataev:1995vh,ChBK:tau:as4nf2}
$$
d_4^{\rm pred}(n_f=3,4,5) = 27 \pm 16, 8 \pm 18,-8 \pm 44
$$
differ significantly from the exact result,  
\beq
d_4^{\rm exact}(n_f=3,4,5) = 49.08,\ \ 27.39,\ \ 9.21
{}.
\eeq
However, within the error estimates \cite{ChBK:tau:as4nf2},  predicted and exact values are  
in agreement. 
%
The picture changes, once these estimates are used to predict the
coefficient $r_4$. Although sizable cancellations between ``dynamical'' and
``kinematical'' terms are observed for the individual $n_f$ coefficients in
(\ref{r_4_separated})
the predictions  for the final results
are significantly closer (in  relative sense) to the results of the exact calculation:
\bea
r_4^{\rm pred}(n_f=3,4,5) &=& 
-129 \pm 16,\   -112 \pm 30, -97 \pm 44
{},
\nonumber
\\
\nonumber
r_4^{\rm exact}(n_f=3,4,5) &=& - 106.88, \ \ \  -92.898, \ \  \  -79.98
{}.
\eea
(This is in striking contrast to the case  of the scalar correlator, where the
predictions for the dynamical terms work well, but, as a consequence of the
strong cancellations between dynamical and kinematical
terms fail in the Minkowskian region \cite{Baikov:2005rw}.) 

Using FAC and the exact result for $d_4$, the
coefficients $d_5$ and $r_5$ can be predicted (following \cite{Kataev:1995vh,ChBK:tau:as4nf2}) 
for $n_f=3,4,5$, namely 
\bea
d_5^{\rm pred}(n_f=3,4,5) &=& 
275,\ \  152,  \ \ 89
{},
\label{d5_pred}
\\
r_5^{\rm pred}(n_f=3,4,5) &=& -505, \ \ -134,\ \  168
\label{r5_pred}
{}.
\eea
These terms may become of relevance for the
International Linear Collider (ILC) running in the GIGA-Z mode with an
anticipated precision of \mbox{$\delta\alpha_s = 0.0005$ --- $0.0007$ \cite{Winter2001}}, and 
already today for the analysis of $\tau$-decays.
 

From the combined analysis of data for 
$ \sigma(e^+e^-\to {\rm hadrons}) $
in the region between 3 and 10 GeV a value
\beq
 \alpha_s(9 \ \mbox{GeV}) = 0.182 \pm 0.033
 \label{alpha_5_GEV}
\eeq
has been obtained recently  \cite{Kuhn:2007tc}.  The shift in
$\alpha_s$ from the inclusion of the $\alpha_s^4$ term amounts to
$\delta \alpha_s(9 \ \mbox{GeV}) = 0.003$ and is thus irrelevant compared to
the large experimental error.

The situation is different for  $Z$ decays. 
The analysis of the electroweak working  group \cite{Alcaraz:2007ri} 
is based on eq.~(\ref{R_numerical_nl})
with  $n_f=5$, including term  up to ${\cal O}(\alpha_s^3)$ and leads to
\beq
 \alpha_s(M_Z)^{NNLO} = 0.1185 \pm 0.0026
 \label{alpha_MZ_old}
\eeq
Since additional corrections (mixed QCD/electroweak or mass terms)
are only weekly  $\alpha_s$-dependent we may consider  
$R(s=M_Z^2)$ as a pseudo-observable:
\beq
R(s=M_Z^2) = 1.03904 \pm 0.00087
 \label{RZ_exp}
{}.
\eeq
Including  the $\alpha_s^4$ term leads to a shift   
$\delta \alpha_s(M_Z) = 0.0005$ 
\beq
 \alpha_s(M_Z)^{NNNLO} =  0.1190 \pm {0.0026}^{\rm exp}
{},
\label{alpha_MZ_new}
\eeq
The theory  error may either be conservatively based on the shift produced  by the last term  (0.0005) or
on the scale variation with $\mu/M_Z = \frac{1}{3} \div 3$,
leading to $\pm 0.0002$  and can be neglected  in both cases.

Higher orders are of particular relevance in the low-energy region, for example
in $\tau$ decays. The correction   from perturbative QCD to the ratio 
\beq
\begin{array}{c}
R_{\tau,V+A} = \frac{\Gamma(\tau\rightarrow{\rm hadrons}_{S=0}
  +\nu_\tau)}{\Gamma(\tau\rightarrow l+\bar\nu_l+\nu_\tau)}
\\
= 3 |V_{ud}|^2 S_{EW}(1+\delta_0 + \delta'_{EW}+ \delta_2+ \delta_{NP})
\end{array}
\label{Rtau1}
\eeq
is given by 
\beq     
1+\delta_0 =
2\int_0^{M_{\tau}^2}\frac{ds}{M_{\tau}^2}
\left(1-\frac{s}{M_{\tau}^2}\right)^2
\left( 1+ \frac{2 s }{M_\tau^2}
\right)
\, R(s)
{}.
\label{equivalent.repr}
\eeq
In the subsequent analysis  we will use 
$S_{EW} = 1.0198 \pm 0.0006$ and $\delta'_{EW} = 0.001$ for the electroweak corrections,
$\delta_2 = (-4.4 \pm 2.0)\times 10^{-4}$ for light quark mass effects,
$\delta_{NP} = (-4.8 \pm 1.7)\times  10^{-3}$ for the nonperturbative effects 
\cite{Schael:2005am,Davier:2005xq}
and $V_{ud} =  0.97418  \pm \, 0.00027$ \cite{PDG}.

The  perturbative quantity $\delta_0 $ can be evaluated in Fixed Order perturbation theory
 or with ``Contour Improvement'' as proposed in \cite{Pivovarov:1991rh,LeDiberder:1992te}
\bea
\delta_0^{FO} &=&   a_s + 5.202 \ a_s^2 +  26.366 \ a_s^3  +127.079 \, a_s^4 
\label{Rtau_FO}
{},
\\
\delta_0^{CI} &=& 
 1.364 \, a_s+ 2.54 \, a_s^2 + 9.71 \, a_s^3
 + 64.29 \, a_s^4
{}.
\label{Rtau_CI}
\eea
(To obtain the $\alpha_s$-dependent  coefficients in  eq.~(\ref{Rtau_CI}) we 
follow \cite{Pivovarov:1991rh,LeDiberder:1992te,Davier:2007ym,Groote:1997cn} and 
use  $\alpha_s(M_{\tau}) = 0.334$ as  reference value.)
For the subsequent analysis  we will use as starting point 
$
\delta_0^{\rm exp}= 0.1998  \pm 0.0043_{\rm exp}
$
as  obtained  from  \cite{Davier:2007ym}
and 
$R_{\tau,V+A} = 3.471 \pm 0.011$
which in turn is based 
on the ''universality-improved'' electronic branching ratio
$B_e= (17.818 \pm 0.032)\%$
and the world  average of the ratio of strange  hadronic and electronic widths 
$0.1686 \pm 0.0047$.
The  new values of $\alpha_s(M_\tau)$ in    
dependence on the choice of $d_4$ (with the previous estimate and the new exact result)
 are summarized in  Table I.

\begin{table}[h]
\caption{\label{tab1}
Results for  $\alpha_s(M_{\tau})$ for different  values of  $d_4$.
The first line displays the $\alpha_s^3$ results (with the $\alpha_s^4$ terms
set to zero in eqs.~(\ref{Rtau_FO},\ref{Rtau_CI})).
The second  line  uses the  the previously predicted  value for $d_4$,
the last  one uses  the exact result (\ref{d4}).     
The first error is the experimental one;
the second (theoretical) uncertainty in the value of $\alpha_s$ corresponds
to changing the  renormalization scale  $\mu$ as follows $\mu^2/M_{\tau}^2 = 0.4 - 2$.
}
\begin{ruledtabular}
\begin{tabular*}{\hsize}{l@{\extracolsep{0ptplus1fil}}c@{\extracolsep{0ptplus1fil}}r}
          
        &
          $\alpha_s^{FO}(M_{\tau})$
        &
          $\alpha_s^{CI}(M_{\tau})$

          \\
        \hline
          -\!\!-
          &
           $0.337 \pm 0.004\ice{_{\rm exp}} \pm 0.03$ 
          &
          $0.354  \pm 0.006\ice{_{\rm exp}} \pm 0.02$ 
          \\
         $d_4 = 25$
          &
        $0.325 \pm 0.004\ice{_{\rm exp}} \pm 0.02$ 
          &
         $0.347 \pm 0.006\ice{_{\rm exp}} \pm 0.009$ 
       \\
       $d_4 = 49.08$      
        &
     $0.322 \pm 0.004\ice{_{\rm exp}} \pm  0.02 $ 
         &
     $0.342  \pm 0.005\ice{_{\rm exp}} \pm  0.01$ 
       \\
\end{tabular*}
\end{ruledtabular}
\end{table}

As stated above the theory error for $\alpha_s$ from Z  decays is small compared to the experimental
uncertainties.  The situation is more problematic for $\tau$ decays and to some extent the 
theory error remains to be  matter of choice.  As anticipated in \cite{ChBK:tau:as4nf2} it
decreases significantly, once $\alpha_s^4$
terms are included. However, the difference between the two methods stabilizes
(this was checked in \cite{ChBK:tau:as4nf2} by adding an estimate for the   $\alpha_s^5$ term) 
and must be considered as irreducible uncertainty. Given the input specified
above we obtain as our final result
\beq
\alpha_s(M_\tau) =0.332 \pm 0.005_{\rm exp}      \pm 0.015_{\rm th} 
{}.
\label{als_tau}
\eeq
For the central value we take  the mean value of FO and CI.
For the theory error we take  half of the difference between
two methods (0.01) plus (module of) the estimated correction from $\alpha_s^5$ term 
(-0.005), the latter being based on $d_5 = 275$ (see eq.~(\ref{d5_pred})).

Applying  four-loop running and matching \cite{vanRitbergen:1997va,Czakon:2004bu,Schroder:2005hy,Chetyrkin:2005ia} 
to (\ref{als_tau}) we
arrive at 
\bea
\label{eq:asres_mz}
   \as(M_Z) &=& 0.1202 \pm 0.0006_{\rm exp}
                                       \pm 0.0018_{\rm th}
                                       \pm 0.0003_{\rm evol}~, \nonumber \\
                     &=& 0.1202  \pm  0.0019 
{}.
\eea
Here the evolution error receives 
contributions from the uncertainties in the $c$-quark mass (0.00003,
$m_c(m_c)=1.286(13)$ GeV \cite{Kuhn:2007vp}) and the $b$-quark mass 
(0.00001, $m_b(m_b)= 4.164(25)$ GeV \cite{Kuhn:2007vp}), 
the matching scale (0.0001, $\mu$ varied between
$0.7\,m_q(m_q)$ and $3.0\,m_q(m_q)$), the four-loop truncation in the
matching expansion (0.0001) and the four-loop truncation in the RGE
equation (0.0003). (For the last two errors the size of the shift due
the highest known perturbative term was treated as systematic
uncertainty.)  The errors are added in quadrature.

Summary: The exact result for the $\alpha_s^4$ term in the Adler
function allows to extract the strong coupling constant from $Z$ and
$\tau$ decays with high precision. Including the exact $\alpha_s^4$
leads to small shifts of the
central values and to a significant reduction of the theory
uncertainty. Note that the shifts in $\alpha_s(M_Z)$
from Z- and $\tau$-decays,
are opposite in sign and move  the values in the  {\em proper} direction,  
decreasing, thus, the current slight mismatch between
two independent  determinations of $\alpha_s$.

The final results 
\beq
\begin{array}{c}
\alpha_s(M_Z)|_{Z}  = 0.1190 \pm 0.0026
\nonumber
{},
\\
\alpha_s(M_Z)|_{\tau }  = 0.1202 \pm 0.0019
\nonumber
\end{array}
\eeq
from these two observables, although based on measurements of
vastly different energy scales, are in remarkable agreement. This constitutes 
a striking test    of asymptotic freedom in QCD. The two values can be combined to
\[
\alpha_s(M_Z) =0.1198 \pm  0.0015
{}.
 \] 
This  is one of the most precise and presently only 
result in order  $\alpha_s^4$.

We thank G.~Quast   for  useful discussions and  M.~Steinhauser 
for reading the manuscript and good advice.
This work was supported by
the Deutsche Forschungsgemeinschaft in the
Sonderforschungsbereich/Transregio
SFB/TR-9 ``Computational Particle Physics'',  by INTAS (grant
03-51-4007) and by RFBR (grants 05-02-17645, 08-02-01451) ).
The computer  calculations were partially  performed on  the  HP XC4000  super
computer of  the  federal state Baden-W\"urttemberg at the High Performance Computing Center Stuttgart 
(HLRS) under the grant ``ParFORM''.

\end{document}